\documentclass[onecolumn,manuscript]{revtex4}
\usepackage{amssymb}

\usepackage{epsfig}
\usepackage[english]{babel}
\usepackage{latexsym}
\usepackage{graphics}
\usepackage{subfigure}
\usepackage{epsfig}
\usepackage{graphicx}
\usepackage{dcolumn}
\usepackage{amsmath}

\newcommand{\be}{\begin{equation}}
\newcommand{\ee}{\end{equation}}
\newcommand{\bey}{\begin{eqnarray}}
\newcommand{\eey}{\end{eqnarray}}
\newcommand{\bw}{\begin{widetext}}
\newcommand{\ew}{\end{widetext}}

\begin{document}                

\title{Pulse-duration dependence of High-order harmonic generation with coherent superposition state}

\author{Bingbing Wang, Taiwang Cheng, Xiaofeng Li, and Panming Fu$^*$}
\address{Laboratory of Optical Physics, Institute of Physics, Chinese Academy of Sciences,
Beijing 100080, China}
\author{Shigang Chen and Jie Liu}
\address{Institute of Applied Physics and Computational
Mathematics, Beijing 100088, China}

\begin{abstract}
We make a systematic study of high-order harmonic generation (HHG)
in a He$^+$-like model ion when the initial states are prepared as
a coherent superposition of the ground state and an excited state.
It is found that, according to the degree of the ionization of the
excited state, the laser intensity can be divided into three
regimes in which HHG spectra exhibit different characteristics.
The pulse-duration dependence of the HHG spectra in these regimes
is studied. We also demonstrate evident advantages of using
coherent superposition state to obtain high conversion efficiency.
The conversion efficiency can be increased further if ultrashort
laser pulses are employed.

\end{abstract}

\par\noindent

\maketitle

\par\noindent
$^*$ Author to whom correspondence should be addressed.

\section{Introduction}

High-order harmonic generation (HHG) is a very useful source of
coherent light in the extreme ultraviolet and soft x-ray regions
of the spectrum [1-4]. HHG occurs when atomic systems interact
with intense laser fields. There are two important aspects we need
to consider in HHG, the cutoff frequency of the harmonic spectrum
and the conversion efficiency of the harmonic generation. The
cutoff frequency is predicted by the cutoff law [5,6], and the
conversion efficiency is decided by the ionization of the atoms.
Many works have been done in increasing the cutoff frequency and
the conversion efficiency, such as by using the ultrashort pulses
[7,8]. Recently, ions have been used to extend the HHG spectrum
cutoff [9-11]. However, the HHG conversion efficiency is usually
very low because the ionization probability is low due to the
large $I_p$. Increasing the harmonic conversion efficiency by
preparing the initial state as a coherent superposition of two
bound states was first proposed by Gauthey et al [12]. Burnett and
co-workers demonstrated that a harmonic spectrum with distinct
plateaus could be obtained by such superposition states. Ishikawa
[11] showed that the conversion efficiency of HHG by He$^+$ ions
can be increased effectively by applying an additional harmonic
pulse to populate one of the excited states. More recently,
Averbukh [13] investigated the atomic polarization effects on HHG
by preparing the initial state as a coherent superposition of one
S state and one P state of atoms. The superposition state can be
obtained by multiphoton resonant excitation [14] or using one
harmonic pulse with the frequency corresponding to the energy
difference between the two bound states [11] before the
fundamental laser pulse.

\par

The idea of preparing the initial state as a coherent
superposition of the ground state and an excited state is that it
can induce dipole transitions between the continuum and the ground
state via the excited state responsible for the ionization. This
process depends on the degree of the ionization of the excited
state. In this paper, we will make a systematic study of HHG with
coherent superposition state in a He$^+$-like model ion. It is
found that, according to the degree of the ionization of the
excited state, the laser intensity can be divided into three
regimes in which HHG spectra exhibit different characteristics.
The pulse-duration dependence of the HHG spectra in these regimes
is studied. We also demonstrate evident advantages of using
coherent superposition state to obtain high conversion efficiency.
The conversion efficiency can be increased further if ultrashort
laser pulses are employed.

\vfill
\section{numerical method}

Our theory is based on solving the one-dimensional time-dependent
Schr\"odinger equation for a hydrogen-like ion in a laser pulse,
which can be expressed as (atomic unit are used throughout):
\begin{eqnarray}
i{\partial \psi (x,t)\over \partial t}=[-{1\over 2}\nabla^2-{a\over \sqrt{b+x^2}}-xE(t)]\psi (x,t),\label{e1}
\end{eqnarray}
where $a$ and $b$ are the parameters describing different ions. We
set $a$=2 and $b$=0.5 in order to get the same ground state
binding energy of He$^+$ ion, i.e. 2.0 a.u., and the second
excited state binding energy is 0.53 a.u in this one-dimension
case. We consider the second excited state rather than the first
excited state because it has the same symmetry of the ground state
and has approximately the same binding energy as the first excited
state of the real He$^+$ ion. $E(t)=F(t) \sin(\omega t+\phi)$ is
the electric field of the pulse. Here, we choose $\omega = 0.056$
(wavelength 800nm) and $\phi=0$ in the calculations. $F(t)$ is the
pulse envelope, which equals $\sin (\pi t /T)^2$ for 10fs pulses,
while
\par\noindent
$F(t) = \left\lbrace
\begin{array}{c l}
    \sin (\pi t/\tau)^2 & \text{if $0<t<\tau/2$},\\
    1 & \text{if $\tau/2<t<T-\tau/2$},\\
    1-\sin (\pi (T-t) /\tau)^2  & \text{if $T-\tau/2<t<T$}.
\end{array}
\right. $
\par\noindent
for 100fs pulses, where $\tau$ is the period of the optical cycle
and $T$ is the laser pulse duration. Equation (1) is integrated
numerically with the help of fast Fourier transform algorithm
[15], where the length of the integration grid is 800, the spatial
step is $dx=0.1$ and the time increment is 0.0125. To avoid
reflections of the wave packet from the boundaries, after each
time increment the wave function is multiplied by a $\cos^{1/8}$
mask function that varies from 1 to 0 over a range from $|x| =300$
to 400.

\section{HHG with coherent superposition state}

The HHG spectrum can be obtained from the Fourier transform of the
time-dependent dipole acceleration
$D(t)=<\psi(x,t)|\nabla|\psi(x,t)>$, which can be written as:
\begin{eqnarray}
D(t)\propto &&  <\psi_{\rm bound}(x,t)|\nabla|\psi_{\rm continuum}(x,t)>
+ c.c..
\end{eqnarray}
Here, we neglect the continuum-continuum transitions because they
have no significant influence to harmonic generation. We prepare
the initial state in a superposition of the ground state $\mid g
\rangle$ and some excited state denoted by $\mid e \rangle$, i.e.,
\begin{eqnarray}
\psi(x,t \rightarrow -\infty)= {1\over \sqrt{2}}(\mid g \rangle +
\mid e \rangle),
\end{eqnarray}
where the phase difference between the states is set to zero for
simplicity. If we assume that the ground and excited states are
not coupled to any other bound state during the pulse, then the
time-dependent wave functions can be written in the form
\begin{eqnarray}
\psi(x,t)=\alpha(t) e^{-i\omega_gt} \mid g \rangle + \beta(t) e^{-i\omega_et}
\mid e \rangle+ \int d k \gamma_k(t) e^{-i\omega_kt} \mid \phi_k(t) \rangle.
\end{eqnarray}
In this expression $|\phi_k(t)>$ is the continuum states
characterized by the momentum $k$, and $\alpha(t)$, $\beta(t)$ and
$\gamma_k(t)$ are the time-dependent amplitudes of the ground,
excited and continuum states, respectively. Here, we have
factorized out the energy dependence of the bare states.
Accordingly, the temporal evolution of the bound state is
\begin{eqnarray}
\psi_{\rm bound}(x,t) = \alpha(t) e^{-i\omega_gt} \mid g \rangle + \beta(t) e^{-i\omega_et}
\mid e \rangle,
\end{eqnarray}
and, we have the time-dependent dipole moment
\begin{eqnarray}
D(t)=D_{gg}(t)+ D_{ee}(t) + D_{ge}(t) + D_{eg}(t),
\end{eqnarray}
where
\begin{eqnarray}
D_{gg}(t)\propto \int d k  \alpha(t) \gamma^g_k(t)
e^{-i(\omega_g-\omega_k)t}<g|\nabla|\phi_k(t)> + c.c.,
\end{eqnarray}
\begin{eqnarray}
D_{ee}(t)\propto \int d k  \beta(t)\gamma^e_k(t)
e^{-i(\omega_e-\omega_k)t}<e|\nabla|\phi_k(t)> + c.c.,
\end{eqnarray}
\begin{eqnarray}
D_{ge}(t)\propto \int d k  \alpha(t) \gamma^e_k(t)
e^{-i(\omega_g-\omega_k)t} <g|\nabla|\phi_k(t)>+ c.c.,
\end{eqnarray}
and
\begin{eqnarray}
D_{eg}(t)\propto \int d k  \beta(t) \gamma^g_k(t)
e^{-i(\omega_e-\omega_k)t} <e|\nabla|\phi_k(t)>  + c.c.,
\end{eqnarray}
where $\gamma^g_k(t)$ ($\gamma^e_k(t)$) is the amplitude of the
continuum state $|\phi_k(t)>$ originated from the ionization of
the ground (excited) state, which by using the strong-field
approximation of Lewenstein et al [16] can be written as[13]
$$\gamma^g_k(t)=i\int^t_0dt'\alpha(t')e
E(t')<k+A(t)/c-A(t')/c|x|g>
exp\{-i\int^t_{t'}{[k+A(t)/c-A(t')/c]^2 \over 2}dt"\},$$ (here
$A(t)$ is the vector potential of the laser pulse). Physically,
$D_{gg}(t)$ and $D_{ee}(t)$ are simply the dipole moments one
would obtain starting in the ground and excited states,
respectively. On the other hand, $D_{ge}(t)$ ($D_{eg}(t)$) can be
regarded as the interference term, where the excited state $\mid e
\rangle$ (the ground state $\mid g \rangle$) is coupled to the
continuum, inducing dipole moments between the continuum and the
ground state $\mid g \rangle$ (the excited state $\mid e
\rangle$). It is important to mention that the tunneling
ionization is usually much easier for electrons at the excite
state than at the ground state. On the other hand, the
probabilities of transitions from the continuum back to the ground
state is higher than that to the excited states. Specifically, as
discussed by Burnett {\sl et al.} [10], we have $\mid
<e|\nabla|\phi_k(t)> \mid / \mid <g|\nabla|\phi_k(t)> \mid\approx
(\omega_g / \omega_e)^{(5/2)}$, which equals approximately 30 in
our case.

\par

We are interested in producing high-energy harmonics photons with
high conversion efficiency. In principle, ions can produce
higher-energy harmonics due to their large ionization potentials
and higher saturation intensities because the cutoff frequency
equals $I_p + 3.2 U_p$. However, harmonics signal for ions has
been shown to be very weak because the efficiency of the harmonic
signal is directly proportional to the ionization rate.  On the
other hand, it is much easier to promote the electron into the
continuum from the excited state. As pointed out by Burnett and
co-workers [10], a possible way of increasing the harmonic
efficiency is to prepare the initial state as a coherent
superposition of the ground state and an excited state so that
dipole transitions are induced between the continuum and the
ground state, where the excited state is responsible for the
ionization (i.e., $D_{ge}(t)$ term in Eq. (9)).

\par

Equations (7)-(10) also indicate that dipole moments are directly
related to the time-dependent amplitudes of the bound states. This
is because harmonic generation originates from the coherent dipole
transition between the continuum and the bound states. As a
result, only those states that remain populated during the pulse
will contribute to the harmonic generation [17].

\section{numerical results}

We will divide the laser intensity into three regimes, according
to the degree of the ionization of the excited state. Figure 1
presents the populations of the ground and second excited states
as a function of time when the initial state is a coherent
superposition of the ground and excited states with equally
weighted populations. The laser pulse duration is 10 fs and
intensity is $I$ = (a) $1\times 10^{13}$ W/cm$^2$, (b) $5\times
10^{14}$ W/cm$^2$ and (c) $4\times 10^{15}$ W/cm$^2$. In the
weak-field regime [Fig. 1(a)] there is only small transference of
population from the excited state to the continuum. In contrast,
the population of the excited state decreases significantly within
the first two optical cycles [from 0.5 to 0.01 within 1.5 optical
cycles in Fig. 1(b)] in the intermediate-field regime; while, in
the strong-field regime the excited state is depleted almost
completely before the peak of the laser pulse [Fig. 1(c)]. Since
ionization plays a crucial role in the generation of harmonics
photons, we will demonstrate that the HHG spectrum shows very
different characteristics in different regimes. Furthermore, by
comparing the HHG spectra for short and long laser pulses, we find
that the spectra exhibit distinct pulse-duration effects,
especially when the laser intensity is high.

\subsection{weak-field regime}

We first study the harmonic generation in the weak-field regime in
which there is only small ionization of the excited state [Fig.
1(a)]. The solid curves in Fig. 2 show the HHG spectra of He$^+$
ion for a coherent superposition state with laser intensity
$1\times 10^{13}$ W/cm$^2$ and pulse duration (a) 10fs and (b)
100fs. For comparison, we also present results when the initial
state is the ground state (dot curve),i.e. $\alpha(0)=1$and
$\beta(0)=0$, and the second excited state (dash curve),
i.e.$\alpha(0)=0$ and $\beta(0)=1$. We should mention that the
harmonic spectra flatten out at the upper end (in figure 2, 3 and
5) is caused by the background numerical noise, has no physical
meaning, and this noise doesn't effect the spectra results. The
HHG spectra of the superposition state (solid curves) clearly
shows two different sets of harmonics. The first one agrees well
with the spectrum of the excited state case (dash curves), while
the second one is about three orders of magnitude higher than that
of the ground state case (dot curve) with the same cutoff harmonic
frequencies.

\par

In the weak-field regime, the amplitudes of the ground and excited
states are approximately constant during the laser pulse (see Fig.
1(a)). On the other hand, it is much easier to ionize the excited
state than the ground state, therefore from Eqs.(7)-(10) we have
$\mid D_{ee}(t) \mid, \mid D_{ge}(t) \mid  \gg \mid D_{gg}(t)
\mid, \mid D_{eg}(t) \mid$. In other words, harmonics of the
superposition case originate from the recombination into the
ground and excited states of electrons, where the excited state is
responsible for the ionization. The maximum kinetic energy that
the electron brings back equals $3.17 U_p$, therefore, when it
recombines into the ground state the energies of the emitted
photons are between $I_g$ and $I_g+ 3.17 U_p$. On the other hand,
recombination into the excited state gives harmonics of energy
between $I_e$ and $I_e+ 3.17 U_p$. The two plateaus will be
separated if $I_g - I_e > 3.17U_p$. In our system the
corresponding laser intensity that the two plateaus can be
separated is lower than $1\times 10^{14}$ W/cm$^2$.

\par

We compare the HHG spectra of the short [Fig. 2(a)] and long [Fig.
2(b)] laser pulses. For long laser pulses, a short burst of
radiation emits every half a laser cycle due to the scatter off
the core of the continuum wave packet. As a result, the
multi-cycle accumulation of the harmonic generation causes
separate sharp peaks in each odd harmonic order. Besides, the
harmonics is usually more intense for long laser pulses,
especially in the first plateau.

\subsection{intermediate-field regime}

Now, let us consider HHG in the intermediate-field regime. Figure
3 presents the harmonic spectra of He$^+$ ion with laser intensity
$5\times 10^{14}$ W/cm$^2$ for the pulse duration (a) 10 fs and
(b) 100 fs. The HHG spectra of the superposition case (solid
curves) show only one plateau, which is about six and five orders
of magnitude higher than that of the ground state case (dotted
curve) when pulse durations are 10 fs and 100 fs, respectively.

\par

As shown in Fig. 1(b) the population of the excited state
decreases significantly within the first two optical cycles in the
intermediate-field regime. Since $\mid \alpha(t) \mid \gg \mid
\beta(t) \mid$ at the time of recombination, we have from Eqs. (8)
and (9) $\mid D_{ge}(t) \mid \gg \mid D_{ee}(t) \mid$. Therefore,
in contrast to the weak-field case, where the recombination into
the ground and  excited states gives two plateaus in the HHG
spectra, the main contribution to the harmonic generation in the
intermediate-field regime is the transition from the continuum to
the ground state. This fact is demonstrated further in Fig. 3
where the HHG spectra of the superposition case (solid curves) is
about three and two orders of magnitude higher than that of the
excited state case (dashed curves) when pulse durations are 10 fs
and 100 fs, respectively.

We are interested in producing harmonic photons with high
conversion efficiency, which is directly proportional to the
population of the continuum and the remain population of the bound
states. In the intermediate-field regime, the laser intensity is
high enough to ionize the excited state within a few optical
cycles, while too weak to directly ionize the ground state.
Therefore, if the initial state is prepared as a coherent
superposition of the ground state and an excited state, a large
dipole transitions will be induced between the continuum and the
ground state, where the excited state is responsible for the
ionization. In our system the intermediate-field regime are from
$I \simeq 1\times 10^{14}$ W/cm$^2$ to about $1\times 10^{15}$
W/cm$^2$. Moreover, Fig. 4 presents the temporal behavior of the
harmonics of the 71th (dashed curve) and 91th (solid curve)
harmonic order for the superposition case when the laser intensity
is $5\times 10^{14}$ W/cm$^2$. It shows that harmonic photons emit
mainly during the first few optical cycles in which the excited
state ionizes efficiently. As a result, conversion efficiency can
be increased further if short laser pulses are employed. For
example, the HHG of 10 fs pulse is on an average one order of
magnitude higher than that of the 100 fs pulse.

\par

Finally, as shown in Fig. 3(b) the HHG spectrum of the excited
state case (dashed curve) exhibits two plateaus with the second
cutoff consistent with that of the ground state case (dotted
curve) when the laser pulse duration is 100 fs. This is because
under the intermediate laser power, there is population transfer
from the excited to the ground states via multiphoton transition.
Therefore, dipole transitions can be induced between the continuum
and the ground state, even the atoms are initially in the excited
state.

\subsection{strong-field regime}

We increase the laser intensity further to a point that there is a
significant population depletion of the excited state within one
optical cycle, and study how this population depletion affects the
HHG spectra. Figure 5 presents the HHG spectra of He$^+$ ion with
$I= 4\times 10^{15}$ W/cm$^2$ for the pulse duration (a) 10 fs and
(b) 100 fs when the initial states are superposition state (solid
curve), ground state (dotted curve) and excited state (dashed
curve).  Let us first consider the HHG spectra with short pulse
duration (Fig. 5(a)). It is found that the spectrum of the ground
state case (dotted curve) exhibits a double-plateau structure. To
understand this, we perform a wavelet time-frequency analysis [18]
of the spectral and temporal structures of HHG. Figure 6(a)
presents the time profile of the harmonics when the initial states
is the ground state. It indicates that the cutoff at about 551th
harmonic emits at time around 1.8 optical cycle. On the other
hand, there are at least four trajectories, which contribute to
the harmonics below the 431th harmonic order, leading to another
plateau with higher strength.

\par

We then consider the excited state case [dashed curve in Fig.
5(a)]. At intensity $I= 4\times 10^{15}$ W/cm$^2$ the excited
state decreases from 0.5 to 0.01 within about 0.7 optical cycles
[Fig. 1(c)]. As a consequence, the cutoff frequency (at about
351th harmonic order) is much smaller than that predicted by the
three-step model, which equals 521th harmonic order according to
the $I_p + 3.17 U_p$ law, because the excited state is depleted
almost completely before the peak of the laser pulse. Moreover, in
contrast to the previous cases, the high depletion of the excited
state also causes the harmonic intensity of the excited state case
much lower than that of the ground state case because $\mid
\alpha(t) \mid \gg \mid \beta(t)\mid$ most of the time [see Eqs.
(7) and (8)].

\par

Now, we consider the spectrum of the superposition case [solid
curve in Fig. 5(a)], which exhibits a complex structure with three
plateaus. The first plateau is about two orders of magnitude
higher than that of the ground state case, while the other part of
the spectrum agrees well with that of the ground state case.
Physically, the HHG spectrum of the superposition case has two
contributions: One originates from the dipole moment $D_{gg}(t)$,
which gives spectrum above the 375th harmonic order and is
consistent with that part of the ground state case. On the other
hand, the first plateau in the spectrum is due to the interference
term $D_{ge}(t)$. The strength of this plateau is about two orders
of magnitude higher than that of the ground state case because of
the large transition from the excited state to the continuum,
demonstrating once again the advantages of using coherent
superposition state to obtain high conversion efficiency. Also,
from the wavelet time-frequency analysis [Fig, 6(b)] we find that
harmonics at the cutoff of this plateau emit at time around 1.4
optical cycle.

\par

Finally, we consider the HHG spectra of the long pulse duration
case [Fig. 5(b)]. First, there is almost no harmonic generation
for the excited state case (dashed curve) because the excited
state is depleted almost completely within one optical cycle.
Second, since there is an effective transition from the ground
state to the continuum while very little depletion of the ground
state population, the conversion efficiency of the ground state
case (dotted curve) is relatively high. Finally, the excited state
plays no role in the harmonic generation when the laser has long
pulse duration, as a result, the HHG spectrum of the superposition
case (solid curve) is consistent with that of the ground state
case. It is worth mentioning that, in the strong-field regime,
there is no advantage of using short pulse. In contrast, the
multi-cycle accumulation causes the conversion efficiency of the
long pulse higher than that of the short pulse by about three
orders of magnitude when the initial state is the ground state.

\section{conclusion}

There are two factors which can affect the conversion efficiency
of HHG, i.e., the ionization rate of the initial bound states and
the remained populations of the bound states at the time of
recombination. The advantage of using coherent superposition state
is that it is possible to induce dipole transitions between the
continuum and the ground state, where the excited state is
responsible for the ionization, thus, drastically increases the
conversion efficiency. In this paper, we make a systematic study
of HHG in a He$^+$-like model ion when the initial states are
prepared as a coherent superposition of the ground state and an
excited state. Since the ionization plays the crucial role in the
HHG with coherent initial state, the laser intensity is divided
into three regimes according to the degree of the ionization of
the excited state. The HHG spectra exhibit different
characteristics in these regimes. We have demonstrated evident
advantages of using coherent superposition state to obtain high
conversion efficiency. We have also found distinct pulse-duration
effects in the intermediate- and strong-field regimes.

This research was supported by the National Natural Science
Foundation of China under Grant No. 60478031, and the Climbing
Programme of the Ministry of Science and Technology of China. B.
Wang thanks Prof. Qiren Zhu, Prof. Armin Scrinzi, Dr. Jing Chen
and Dr. Jiangbin Gong for helpful discussions.

\begin{figure}
\includegraphics[width=0.7\columnwidth]{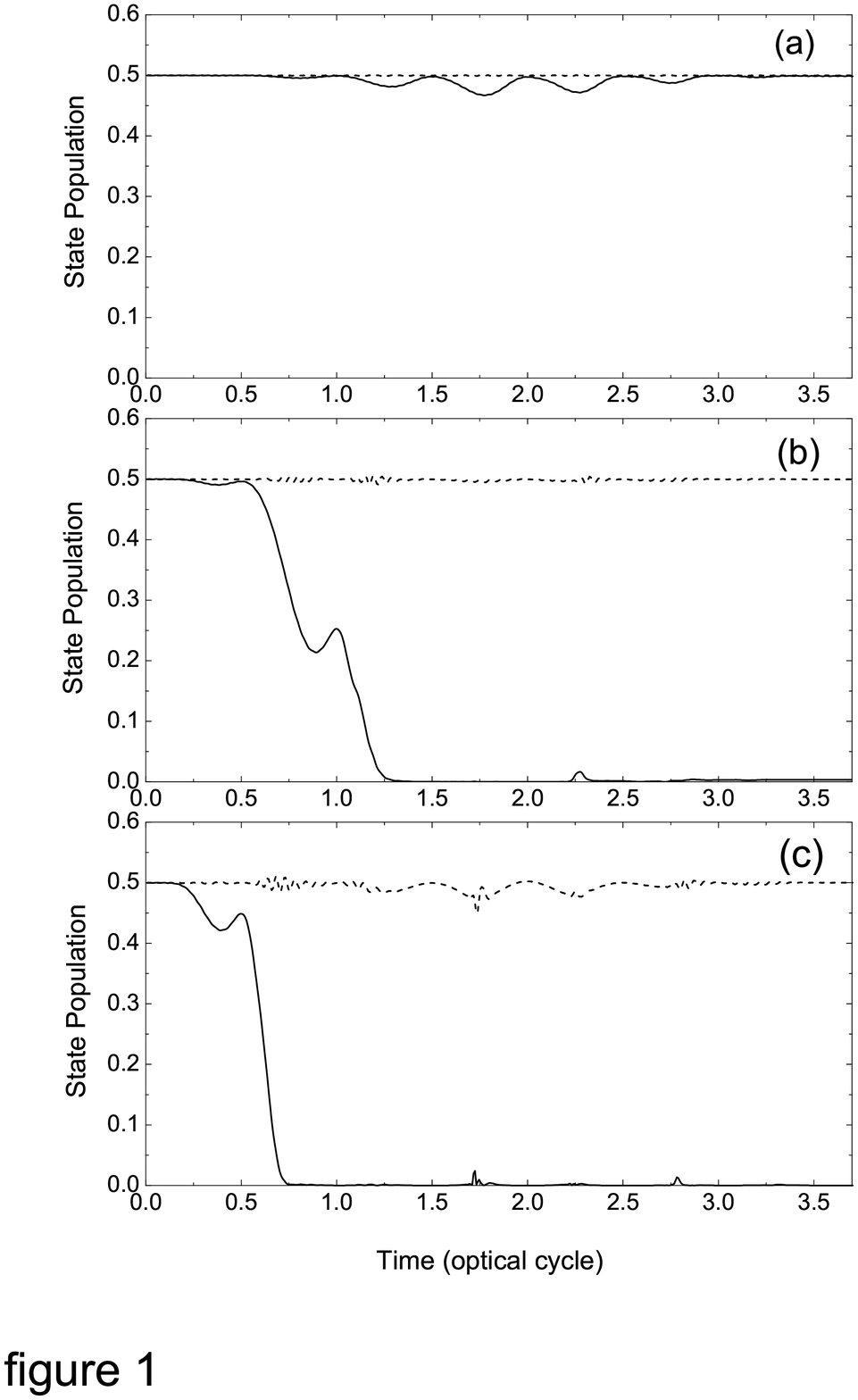} \vspace{-0.2cm}
\caption{Populations of the ground and second excited states as a
function of time when the initial state is a coherent
superposition of the ground and excited states with equally
weighted populations. The laser pulse duration is 10 fs and
intensity is $I$ = (a) $1\times 10^{13}$ W/cm$^2$, (b) $5\times
10^{14}$ W/cm$^2$ and (c) $4\times 10^{15}$ W/cm$^2$.}
\end{figure}
\begin{figure}
\includegraphics[width=0.7\columnwidth]{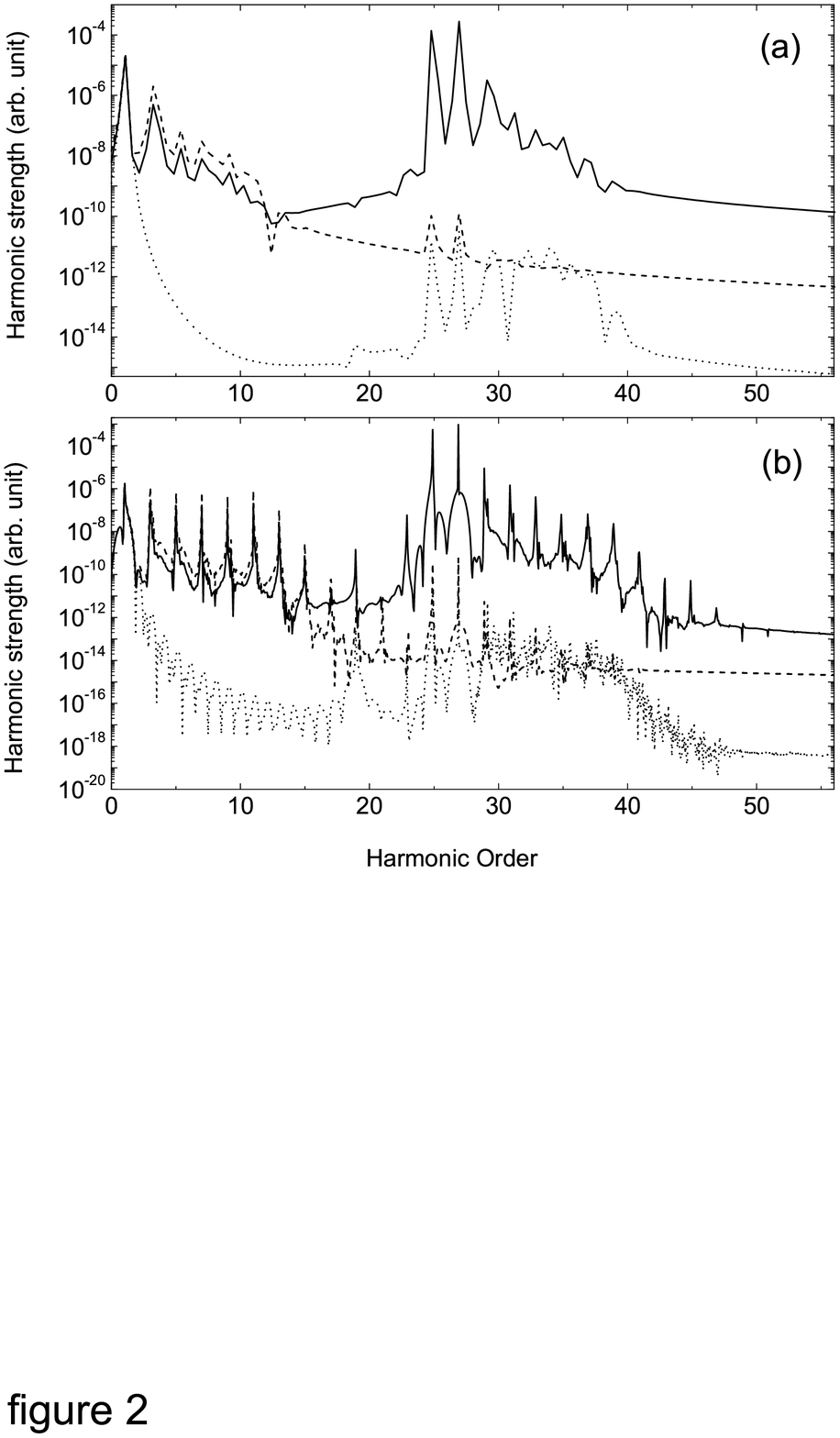} \vspace{-0.2cm}
\caption{Harmonic spectra of He$^+$ ion with laser intensity
$1\times 10^{13}$ W/cm$^2$ and pulse duration (a) 10 fs and (b)
100 fs when the initial states are superposition state (solid
curve), ground state (dotted curve) and excited state (dashed
curve).}
\end{figure}
\begin{figure}
\includegraphics[width=0.7\columnwidth]{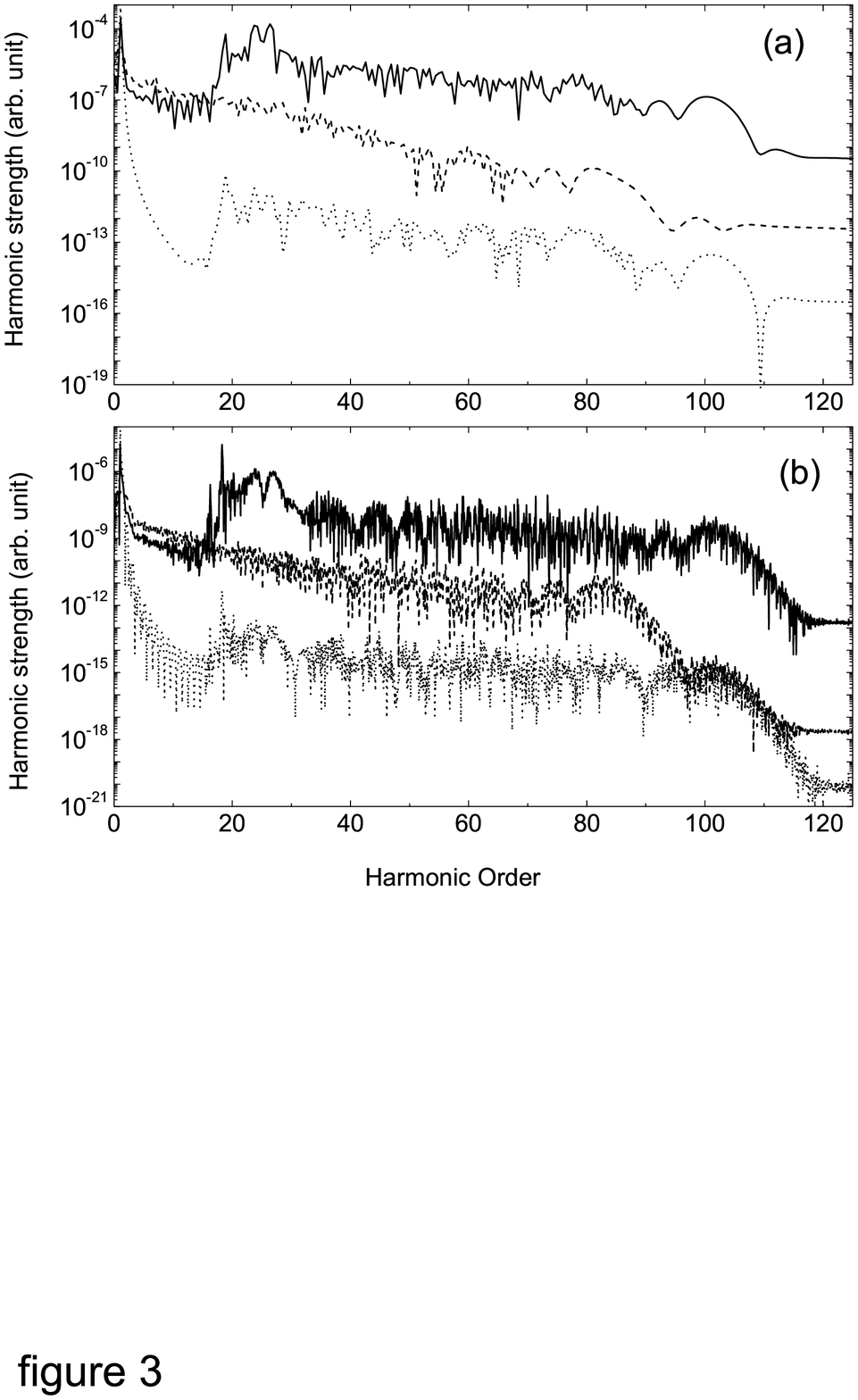} \vspace{-0.2cm}
\caption{Harmonic spectra of He$^+$ ion with laser intensity
$5\times 10^{14}$ W/cm$^2$ and pulse duration (a) 10 fs and (b)
100 fs when the initial states are superposition state (solid
curve), ground state (dotted curve) and excited state (dashed
curve).}
\end{figure}
\begin{figure}
\includegraphics[width=0.7\columnwidth]{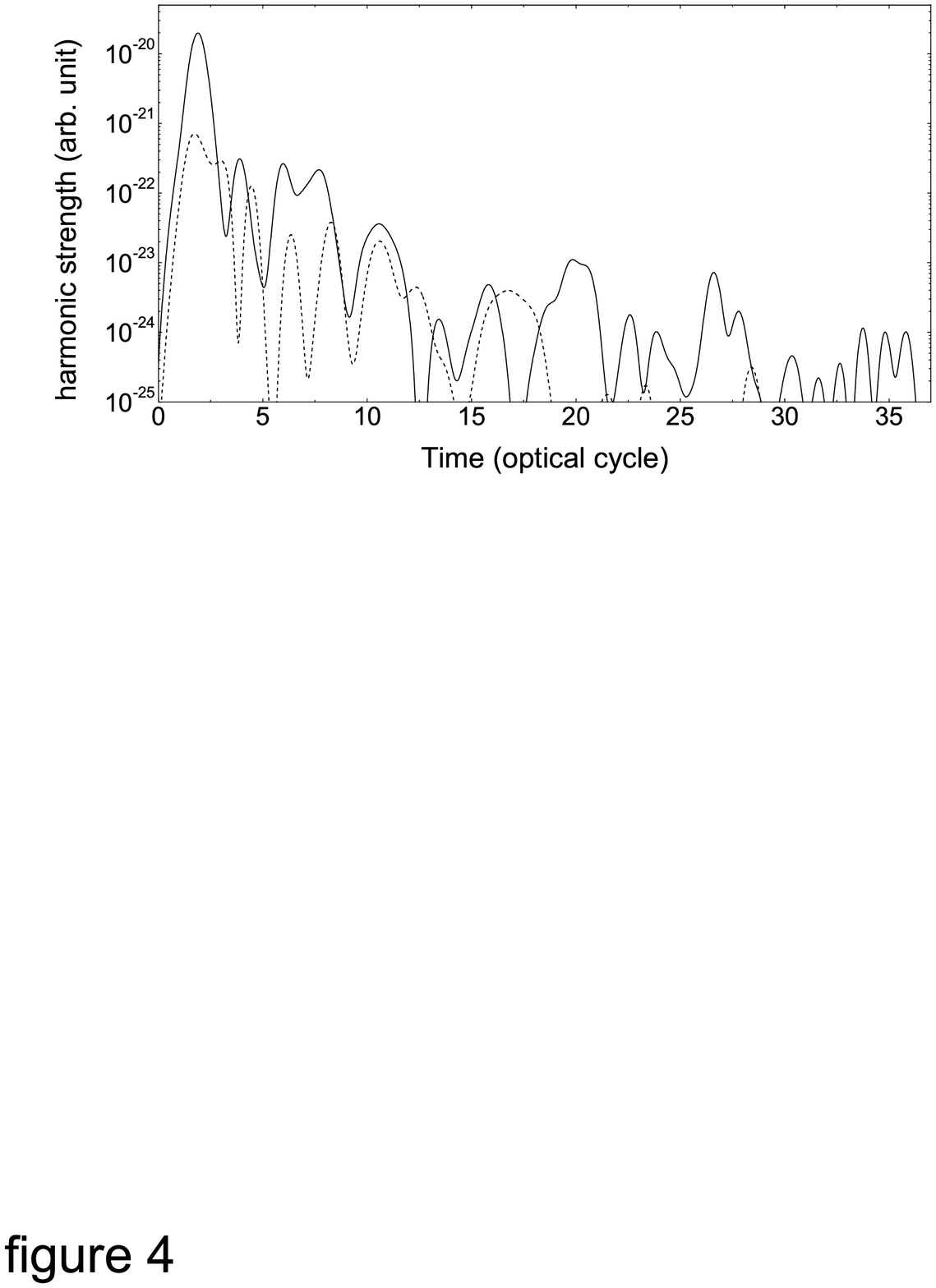} \vspace{-0.2cm}
\caption{Temporal behavior of the harmonics of the 71th (dashed
curve) and 91th (solid curve) harmonic order for the superposition
case when the laser intensity is $5\times 10^{14}$ W/cm$^2$.}
\end{figure}
\begin{figure}
\includegraphics[width=0.7\columnwidth]{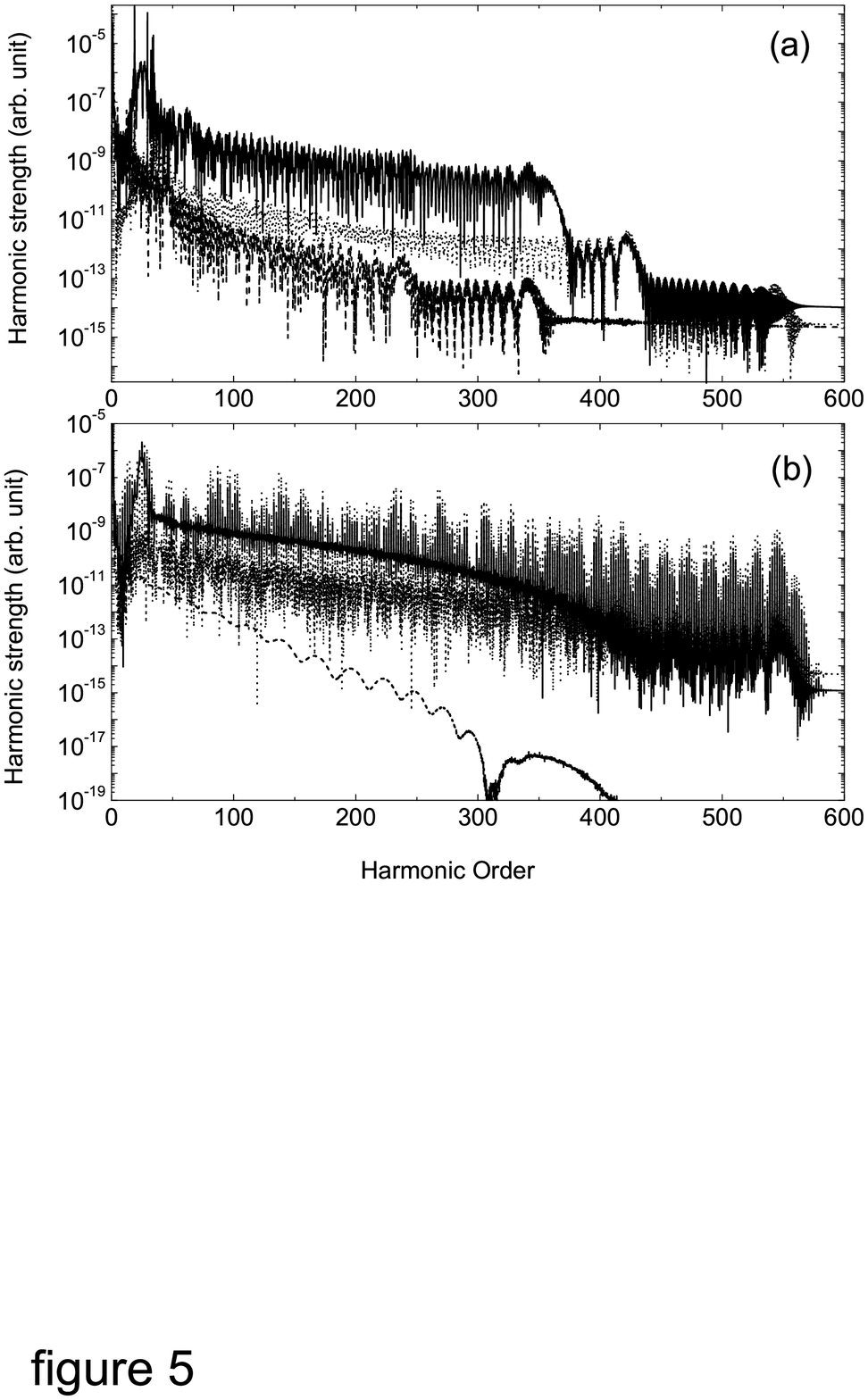} \vspace{-0.2cm}
\caption{Harmonic spectra of He$^+$ ion with laser intensity
$4\times 10^{15}$ W/cm$^2$ and pulse duration (a) 10 fs and (b)
100 fs when the initial states are superposition state (solid
curve), ground state (dotted curve) and excited state (dashed
curve).}
\end{figure}

\begin{figure}
\includegraphics[width= \columnwidth]{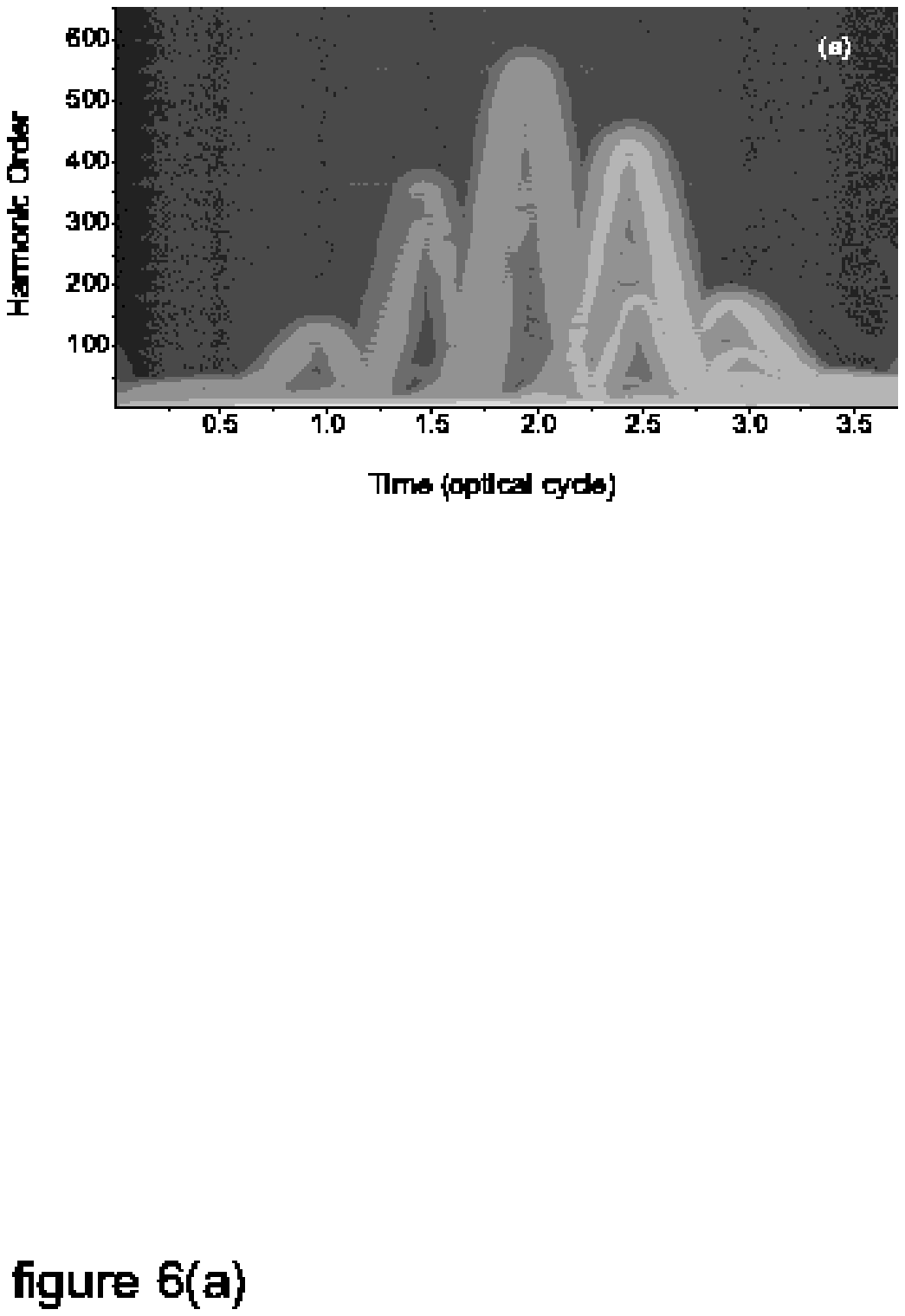}\vspace{-0.2cm}
\end{figure}

\begin{figure}
\includegraphics[width= \columnwidth]{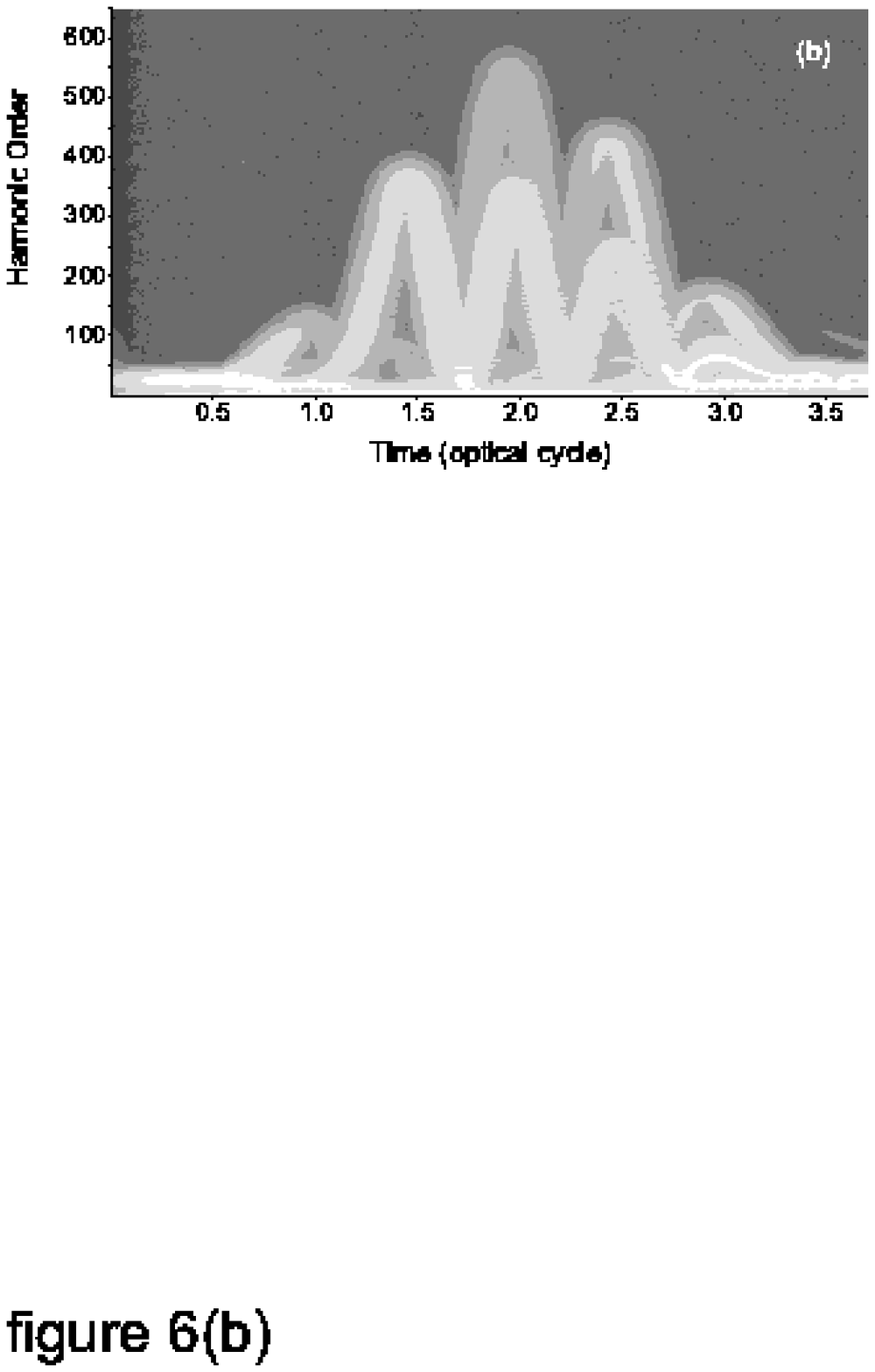}\vspace{-0.2cm}
\caption{Time profile of the harmonics when the initial state is
(a) the ground state and (b) the coherent superposition state.}
\end{figure}

\end{document}